# Observation of the $\Upsilon(1^3D_J)$ bottomonium state through decays to $\pi^+\pi^-\Upsilon(1S)$

P. del Amo Sanchez,[1] J. P. Lees,[1] V. Poireau,[1] E. Prencipe,[1] V. Tisserand,[1] J. Garra Tico,[2] E. Grauges,[2] M. Martinelli[ab,3] A. Palano[ab,3] M. Pappagallo[ab,3] G. Eigen,[4] B. Stugu,[4] L. Sun,[4] M. Battaglia,[5] D. N. Brown,[5] B. Hooberman,[5] L. T. Kerth,[5] Yu. G. Kolomensky,[5] G. Lynch,[5] I. L. Osipenkov,[5] T. Tanabe,[5] C. M. Hawkes,[6] A. T. Watson,[6] H. Koch,[7] T. Schroeder,[7] D. J. Asgeirsson,[8] C. Hearty,[8] T. S. Mattison,[8] J. A. McKenna,[8] A. Khan,[9] A. Randle-Conde,[9] V. E. Blinov,[10] A. R. Buzykaev,[10] V. P. Druzhinin,[10] V. B. Golubev,[10] A. P. Onuchin,[10] S. I. Serednyakov,[10] Yu. I. Skovpen,[10] E. P. Solodov,[10] K. Yu. Todyshev,[10] A. N. Yushkov,[10] M. Bondioli,[11] S. Curry,[11] D. Kirkby,[11] A. J. Lankford,[11] M. Mandelkern,[11] E. C. Martin,[11] D. P. Stoker,[11] H. Atmacan,[12] J. W. Gary,[12] F. Liu,[12] O. Long,[12] G. M. Vitug,[12] C. Campagnari,[13] T. M. Hong,[13] D. Kovalskyi,[13] J. D. Richman,[13] A. M. Eisner,[14] C. A. Heusch,[14] J. Kroseberg,[14] W. S. Lockman,[14] A. J. Martinez,[14] T. Schalk,[14] B. A. Schumm,[14] A. Seiden,[14] L. O. Winstrom,[14] C. H. Cheng,[15] D. A. Doll,[15] B. Echenard,[15] D. G. Hitlin,[15] P. Ongmongkolkul,[15] F. C. Porter,[15] A. Y. Rakitin,[15] R. Andreassen,[16] M. S. Dubrovin,[16] G. Mancinelli,[16] B. T. Meadows,[16] M. D. Sokoloff,[16] P. C. Bloom,[17] W. T. Ford,[17] A. Gaz,[17] J. F. Hirschauer,[17] M. Nagel,[17] U. Nauenberg,[17] J. G. Smith,[17] S. R. Wagner,[17] R. Ayad,[18,*] W. H. Toki,[18] T. M. Karbach,[19] J. Merkel,[19] A. Petzold,[19] B. Spaan,[19] K. Wacker,[19] M. J. Kobel,[20] K. R. Schubert,[20] R. Schwierz,[20] D. Bernard,[21] M. Verderi,[21] P. J. Clark,[22] S. Playfer,[22] J. E. Watson,[22] M. Andreotti[ab,23] D. Bettoni[a,23] C. Bozzi[a,23] R. Calabrese[ab,23] A. Cecchi[ab,23] G. Cibinetto[ab,23] E. Fioravanti[ab,23] P. Franchini[ab,23] E. Luppi[ab,23] M. Munerato[ab,23] M. Negrini[ab,23] A. Petrella[ab,23] L. Piemontese[a,23] R. Baldini-Ferroli,[24] A. Calcaterra,[24] R. de Sangro,[24] G. Finocchiaro,[24] M. Nicolaci,[24] S. Pacetti,[24] P. Patteri,[24] I. M. Peruzzi,[24,†] M. Piccolo,[24] M. Rama,[24] A. Zallo,[24] R. Contri[ab,25] E. Guido[ab,25] M. Lo Vetere[ab,25] M. R. Monge[ab,25] S. Passaggio[a,25] C. Patrignani[ab,25] E. Robutti[a,25] S. Tosi[ab,25] B. Bhuyan,[26] M. Morii,[27] A. Adametz,[28] J. Marks,[28] S. Schenk,[28] U. Uwer,[28] F. U. Bernlochner,[29] H. M. Lacker,[29] T. Lueck,[29] A. Volk,[29] P. D. Dauncey,[30] M. Tibbetts,[30] P. K. Behera,[31] U. Mallik,[31] C. Chen,[32] J. Cochran,[32] H. B. Crawley,[32] L. Dong,[32] W. T. Meyer,[32] S. Prell,[32] E. I. Rosenberg,[32] A. E. Rubin,[32] Y. Y. Gao,[33] A. V. Gritsan,[33] Z. J. Guo,[33] N. Arnaud,[34] M. Davier,[34] D. Derkach,[34] J. Firmino da Costa,[34] G. Grosdidier,[34] F. Le Diberder,[34] A. M. Lutz,[34] B. Malaescu,[34] A. Perez,[34] P. Roudeau,[34] M. H. Schune,[34] J. Serrano,[34] V. Sordini,[34,‡] A. Stocchi,[34] L. Wang,[34] G. Wormser,[34] D. J. Lange,[35] D. M. Wright,[35] I. Bingham,[36] J. P. Burke,[36] C. A. Chavez,[36] J. P. Coleman,[36] J. R. Fry,[36] E. Gabathuler,[36] R. Gamet,[36] D. E. Hutchcroft,[36] D. J. Payne,[36] C. Touramanis,[36] A. J. Bevan,[37] F. Di Lodovico,[37] R. Sacco,[37] M. Sigamani,[37] G. Cowan,[38] S. Paramesvaran,[38] A. C. Wren,[38] D. N. Brown,[39] C. L. Davis,[39] A. G. Denig,[40] M. Fritsch,[40] W. Gradl,[40] A. Hafner,[40] K. E. Alwyn,[41] D. Bailey,[41] R. J. Barlow,[41] G. Jackson,[41] G. D. Lafferty,[41] T. J. West,[41] J. Anderson,[42] R. Cenci,[42] A. Jawahery,[42] D. A. Roberts,[42] G. Simi,[42] J. M. Tuggle,[42] C. Dallapiccola,[43] E. Salvati,[43] R. Cowan,[44] D. Dujmic,[44] P. H. Fisher,[44] G. Sciolla,[44] M. Zhao,[44] D. Lindemann,[45] P. M. Patel,[45] S. H. Robertson,[45] M. Schram,[45] P. Biassoni[ab,46] A. Lazzaro[ab,46] V. Lombardo[a,46] F. Palombo[ab,46] S. Stracka[ab,46] L. Cremaldi,[47] R. Godang,[47,§] R. Kroeger,[47] P. Sonnek,[47] D. J. Summers,[47] H. W. Zhao,[47] X. Nguyen,[48] M. Simard,[48] P. Taras,[48] G. De Nardo[ab,49] D. Monorchio[ab,49] G. Onorato[ab,49] C. Sciacca[ab,49] G. Raven,[50] H. L. Snoek,[50] C. P. Jessop,[51] K. J. Knoepfel,[51] J. M. LoSecco,[51] W. F. Wang,[51] L. A. Corwin,[52] K. Honscheid,[52] R. Kass,[52] J. P. Morris,[52] A. M. Rahimi,[52] N. L. Blount,[53] J. Brau,[53] R. Frey,[53] O. Igonkina,[53] J. A. Kolb,[53] R. Rahmat,[53] N. B. Sinev,[53] D. Strom,[53] J. Strube,[53] E. Torrence,[53] G. Castelli[ab,54] E. Feltresi[ab,54] N. Gagliardi[ab,54] M. Margoni[ab,54] M. Morandin[a,54] M. Posocco[a,54] M. Rotondo[a,54] F. Simonetto[ab,54] R. Stroili[ab,54] E. Ben-Haim,[55] G. R. Bonneaud,[55] H. Briand,[55] G. Calderini,[55] J. Chauveau,[55] O. Hamon,[55] Ph. Leruste,[55] G. Marchiori,[55] J. Ocariz,[55] J. Prendki,[55] S. Sitt,[55] M. Biasini[ab,56] E. Manoni[ab,56] C. Angelini[ab,57] G. Batignani[ab,57] S. Bettarini[ab,57] M. Carpinelli[ab,57,¶] G. Casarosa[ab,57] A. Cervelli[ab,57] F. Forti[ab,57] M. A. Giorgi[ab,57] A. Lusiani[ac,57] N. Neri[ab,57] E. Paoloni[ab,57] G. Rizzo[ab,57] J. J. Walsh[a,57] D. Lopes Pegna,[58] C. Lu,[58] J. Olsen,[58] A. J. S. Smith,[58] A. V. Telnov,[58] F. Anulli[a,59] E. Baracchini[ab,59] G. Cavoto[a,59] R. Faccini[ab,59] F. Ferrarotto[a,59] F. Ferroni[ab,59] M. Gaspero[ab,59] L. Li Gioi[a,59]

M. A. Mazzoni[a],[59] G. Piredda[a],[59] F. Renga[ab],[59] M. Ebert,[60] T. Hartmann,[60] T. Leddig,[60] H. Schröder,[60] R. Waldi,[60] T. Adye,[61] B. Franek,[61] E. O. Olaiya,[61] F. F. Wilson,[61] S. Emery,[62] G. Hamel de Monchenault,[62] G. Vasseur,[62] Ch. Yèche,[62] M. Zito,[62] M. T. Allen,[63] D. Aston,[63] D. J. Bard,[63] R. Bartoldus,[63] J. F. Benitez,[63] C. Cartaro,[63] M. R. Convery,[63] J. Dorfan,[63] G. P. Dubois-Felsmann,[63] W. Dunwoodie,[63] R. C. Field,[63] M. Franco Sevilla,[63] B. G. Fulsom,[63] A. M. Gabareen,[63] M. T. Graham,[63] P. Grenier,[63] C. Hast,[63] W. R. Innes,[63] M. H. Kelsey,[63] H. Kim,[63] P. Kim,[63] M. L. Kocian,[63] D. W. G. S. Leith,[63] S. Li,[63] B. Lindquist,[63] S. Luitz,[63] V. Luth,[63] H. L. Lynch,[63] D. B. MacFarlane,[63] H. Marsiske,[63] D. R. Muller,[63] H. Neal,[63] S. Nelson,[63] C. P. O'Grady,[63] I. Ofte,[63] M. Perl,[63] T. Pulliam,[63] B. N. Ratcliff,[63] A. Roodman,[63] A. A. Salnikov,[63] V. Santoro,[63] R. H. Schindler,[63] J. Schwiening,[63] A. Snyder,[63] D. Su,[63] M. K. Sullivan,[63] S. Sun,[63] K. Suzuki,[63] J. M. Thompson,[63] J. Va'vra,[63] A. P. Wagner,[63] M. Weaver,[63] C. A. West,[63] W. J. Wisniewski,[63] M. Wittgen,[63] D. H. Wright,[63] H. W. Wulsin,[63] A. K. Yarritu,[63] C. C. Young,[63] V. Ziegler,[63] X. R. Chen,[64] W. Park,[64] M. V. Purohit,[64] R. M. White,[64] J. R. Wilson,[64] S. J. Sekula,[65] M. Bellis,[66] P. R. Burchat,[66] A. J. Edwards,[66] T. S. Miyashita,[66] S. Ahmed,[67] M. S. Alam,[67] J. A. Ernst,[67] B. Pan,[67] M. A. Saeed,[67] S. B. Zain,[67] N. Guttman,[68] A. Soffer,[68] P. Lund,[69] S. M. Spanier,[69] R. Eckmann,[70] J. L. Ritchie,[70] A. M. Ruland,[70] C. J. Schilling,[70] R. F. Schwitters,[70] B. C. Wray,[70] J. M. Izen,[71] X. C. Lou,[71] F. Bianchi[ab],[72] D. Gamba[ab],[72] M. Pelliccioni[ab],[72] M. Bomben[ab],[73] L. Lanceri[ab],[73] L. Vitale[ab],[73] N. Lopez-March,[74] F. Martinez-Vidal,[74] D. A. Milanes,[74] A. Oyanguren,[74] J. Albert,[75] Sw. Banerjee,[75] H. H. F. Choi,[75] K. Hamano,[75] G. J. King,[75] R. Kowalewski,[75] M. J. Lewczuk,[75] I. M. Nugent,[75] J. M. Roney,[75] R. J. Sobie,[75] T. J. Gershon,[76] P. F. Harrison,[76] J. Ilic,[76] T. E. Latham,[76] E. M. T. Puccio,[76] H. R. Band,[77] X. Chen,[77] S. Dasu,[77] K. T. Flood,[77] Y. Pan,[77] R. Prepost,[77] C. O. Vuosalo,[77] and S. L. Wu[77]

(The *BABAR* Collaboration)

[1]*Laboratoire d'Annecy-le-Vieux de Physique des Particules (LAPP), Université de Savoie, CNRS/IN2P3, F-74941 Annecy-Le-Vieux, France*

[2]*Universitat de Barcelona, Facultat de Fisica, Departament ECM, E-08028 Barcelona, Spain*

[3]*INFN Sezione di Bari[a]; Dipartimento di Fisica, Università di Bari[b], I-70126 Bari, Italy*

[4]*University of Bergen, Institute of Physics, N-5007 Bergen, Norway*

[5]*Lawrence Berkeley National Laboratory and University of California, Berkeley, California 94720, USA*

[6]*University of Birmingham, Birmingham, B15 2TT, United Kingdom*

[7]*Ruhr Universität Bochum, Institut für Experimentalphysik 1, D-44780 Bochum, Germany*

[8]*University of British Columbia, Vancouver, British Columbia, Canada V6T 1Z1*

[9]*Brunel University, Uxbridge, Middlesex UB8 3PH, United Kingdom*

[10]*Budker Institute of Nuclear Physics, Novosibirsk 630090, Russia*

[11]*University of California at Irvine, Irvine, California 92697, USA*

[12]*University of California at Riverside, Riverside, California 92521, USA*

[13]*University of California at Santa Barbara, Santa Barbara, California 93106, USA*

[14]*University of California at Santa Cruz, Institute for Particle Physics, Santa Cruz, California 95064, USA*

[15]*California Institute of Technology, Pasadena, California 91125, USA*

[16]*University of Cincinnati, Cincinnati, Ohio 45221, USA*

[17]*University of Colorado, Boulder, Colorado 80309, USA*

[18]*Colorado State University, Fort Collins, Colorado 80523, USA*

[19]*Technische Universität Dortmund, Fakultät Physik, D-44221 Dortmund, Germany*

[20]*Technische Universität Dresden, Institut für Kern- und Teilchenphysik, D-01062 Dresden, Germany*

[21]*Laboratoire Leprince-Ringuet, CNRS/IN2P3, Ecole Polytechnique, F-91128 Palaiseau, France*

[22]*University of Edinburgh, Edinburgh EH9 3JZ, United Kingdom*

[23]*INFN Sezione di Ferrara[a]; Dipartimento di Fisica, Università di Ferrara[b], I-44100 Ferrara, Italy*

[24]*INFN Laboratori Nazionali di Frascati, I-00044 Frascati, Italy*

[25]*INFN Sezione di Genova[a]; Dipartimento di Fisica, Università di Genova[b], I-16146 Genova, Italy*

[26]*Indian Institute of Technology Guwahati, Guwahati, Assam, 781 039, India*

[27]*Harvard University, Cambridge, Massachusetts 02138, USA*

[28]*Universität Heidelberg, Physikalisches Institut, Philosophenweg 12, D-69120 Heidelberg, Germany*

[29]*Humboldt-Universität zu Berlin, Institut für Physik, Newtonstr. 15, D-12489 Berlin, Germany*

[30]*Imperial College London, London, SW7 2AZ, United Kingdom*

[31]University of Iowa, Iowa City, Iowa 52242, USA
[32]Iowa State University, Ames, Iowa 50011-3160, USA
[33]Johns Hopkins University, Baltimore, Maryland 21218, USA
[34]Laboratoire de l'Accélérateur Linéaire, IN2P3/CNRS et Université Paris-Sud 11, Centre Scientifique d'Orsay, B. P. 34, F-91898 Orsay Cedex, France
[35]Lawrence Livermore National Laboratory, Livermore, California 94550, USA
[36]University of Liverpool, Liverpool L69 7ZE, United Kingdom
[37]Queen Mary, University of London, London, E1 4NS, United Kingdom
[38]University of London, Royal Holloway and Bedford New College, Egham, Surrey TW20 0EX, United Kingdom
[39]University of Louisville, Louisville, Kentucky 40292, USA
[40]Johannes Gutenberg-Universität Mainz, Institut für Kernphysik, D-55099 Mainz, Germany
[41]University of Manchester, Manchester M13 9PL, United Kingdom
[42]University of Maryland, College Park, Maryland 20742, USA
[43]University of Massachusetts, Amherst, Massachusetts 01003, USA
[44]Massachusetts Institute of Technology, Laboratory for Nuclear Science, Cambridge, Massachusetts 02139, USA
[45]McGill University, Montréal, Québec, Canada H3A 2T8
[46]INFN Sezione di Milano[a]; Dipartimento di Fisica, Università di Milano[b], I-20133 Milano, Italy
[47]University of Mississippi, University, Mississippi 38677, USA
[48]Université de Montréal, Physique des Particules, Montréal, Québec, Canada H3C 3J7
[49]INFN Sezione di Napoli[a]; Dipartimento di Scienze Fisiche, Università di Napoli Federico II[b], I-80126 Napoli, Italy
[50]NIKHEF, National Institute for Nuclear Physics and High Energy Physics, NL-1009 DB Amsterdam, The Netherlands
[51]University of Notre Dame, Notre Dame, Indiana 46556, USA
[52]Ohio State University, Columbus, Ohio 43210, USA
[53]University of Oregon, Eugene, Oregon 97403, USA
[54]INFN Sezione di Padova[a]; Dipartimento di Fisica, Università di Padova[b], I-35131 Padova, Italy
[55]Laboratoire de Physique Nucléaire et de Hautes Energies, IN2P3/CNRS, Université Pierre et Marie Curie-Paris6, Université Denis Diderot-Paris7, F-75252 Paris, France
[56]INFN Sezione di Perugia[a]; Dipartimento di Fisica, Università di Perugia[b], I-06100 Perugia, Italy
[57]INFN Sezione di Pisa[a]; Dipartimento di Fisica, Università di Pisa[b]; Scuola Normale Superiore di Pisa[c], I-56127 Pisa, Italy
[58]Princeton University, Princeton, New Jersey 08544, USA
[59]INFN Sezione di Roma[a]; Dipartimento di Fisica, Università di Roma La Sapienza[b], I-00185 Roma, Italy
[60]Universität Rostock, D-18051 Rostock, Germany
[61]Rutherford Appleton Laboratory, Chilton, Didcot, Oxon, OX11 0QX, United Kingdom
[62]CEA, Irfu, SPP, Centre de Saclay, F-91191 Gif-sur-Yvette, France
[63]SLAC National Accelerator Laboratory, Stanford, California 94309 USA
[64]University of South Carolina, Columbia, South Carolina 29208, USA
[65]Southern Methodist University, Dallas, Texas 75275, USA
[66]Stanford University, Stanford, California 94305-4060, USA
[67]State University of New York, Albany, New York 12222, USA
[68]Tel Aviv University, School of Physics and Astronomy, Tel Aviv, 69978, Israel
[69]University of Tennessee, Knoxville, Tennessee 37996, USA
[70]University of Texas at Austin, Austin, Texas 78712, USA
[71]University of Texas at Dallas, Richardson, Texas 75083, USA
[72]INFN Sezione di Torino[a]; Dipartimento di Fisica Sperimentale, Università di Torino[b], I-10125 Torino, Italy
[73]INFN Sezione di Trieste[a]; Dipartimento di Fisica, Università di Trieste[b], I-34127 Trieste, Italy
[74]IFIC, Universitat de Valencia-CSIC, E-46071 Valencia, Spain
[75]University of Victoria, Victoria, British Columbia, Canada V8W 3P6
[76]Department of Physics, University of Warwick, Coventry CV4 7AL, United Kingdom
[77]University of Wisconsin, Madison, Wisconsin 53706, USA

Based on $122\times10^6$ $\Upsilon(3S)$ events collected with the *BABAR* detector, we have observed the $\Upsilon(1^3D_J)$ bottomonium state through the $\Upsilon(3S) \to \gamma\gamma\Upsilon(1^3D_J) \to \gamma\gamma\pi^+\pi^-\Upsilon(1S)$ decay chain. The significance for the $J = 2$ member of the $\Upsilon(1^3D_J)$ triplet is 5.8 standard deviations including systematic uncertainties. The mass of the $J = 2$ state is determined to be $10164.5 \pm 0.8\,(\text{stat.}) \pm 0.5\,(\text{syst.})$ MeV/$c^2$. We use the $\pi^+\pi^-$ invariant mass distribution to confirm the consistency of the observed state with the orbital angular momentum assignment of the $\Upsilon(1^3D_J)$.

Heavy quark bound states below open flavor thresholds provide a key probe of the interactions between quarks. The mass spectrum and branching fractions of these states can be described by potential models and quantum chromodynamics [1–3]. $S$-wave and $P$-wave bottomonium ($b\bar{b}$) states were first observed in the 1970s and 1980s. Only recently [4] has a $D$-wave bottomonium state, the triplet $\Upsilon(1^3D_J)$ [5], been observed, where $J = 1, 2, 3$. The separation between the members of the triplet (intrinsic widths about 30 keV/$c^2$) is expected to be on the order of 10 MeV/$c^2$ [2]. A single state, interpreted to be the $J = 2$ member of the $\Upsilon(1^3D_J)$ triplet, was observed [4] by the CLEO Collaboration in the radiative $\Upsilon(1^3D_2) \to \gamma\gamma\Upsilon(1S)$ decay channel, but the quantum numbers $L$, $J$ [5] and parity $P$ were not verified.

In this paper, we report the observation of the $J = 2$ state of the $\Upsilon(1^3D_J)$ in the hadronic $\pi^+\pi^-\Upsilon(1S)$ decay channel, with $\Upsilon(1S) \to \ell^+\ell^-$ ($\ell = e$, $\mu$). This decay channel has been of interest for decades [2, 6–8]. Predictions for the branching fraction vary widely [6–8]. It provides better mass resolution than the $\gamma\gamma\Upsilon(1S)$ channel and allows $L$, $J$, and $P$, for which there is currently no experimental information, to be tested, through measurement of the angular distributions of the $\pi^\pm$ and $\ell^\pm$. The only previous result for this channel is the 90% confidence level (CL) branching fraction upper limit $\mathcal{B}_{\Upsilon(3S)\to\gamma\gamma\Upsilon(1^3D_J)} \times \mathcal{B}_{\Upsilon(1^3D_J)\to\pi^+\pi^-\Upsilon(1S)} \times \mathcal{B}_{\Upsilon(1S)\to\ell^+\ell^-} < 6.6 \times 10^{-6}$ [4].

The analysis is based on a sample of $(121.8\pm1.2)\times10^6$ $\Upsilon(3S)$ decays collected with the *BABAR* detector at the PEP-II asymmetric-energy $e^+e^-$ storage rings at the SLAC National Accelerator Laboratory, corresponding to an integrated luminosity of 28.6 fb$^{-1}$. The *BABAR* detector is described elsewhere [9]. Monte Carlo (MC) event samples that include simulation of the detector response are used to determine the signal and background characteristics, optimize selection criteria, and evaluate efficiencies. Pure electric-dipole transitions [10] are assumed when generating radiative decays.

The $\Upsilon(1^3D_J)$ in our study are produced through $\Upsilon(3S) \to \gamma\chi_{bJ'}(2P) \to \gamma\gamma\Upsilon(1^3D_J)$ transitions, with $J' = 0, 1, 2$. To reconstruct the $\Upsilon(3S) \to \gamma\gamma\pi^+\pi^-\ell^+\ell^-$ final states, we require exactly four charged tracks in an event, two of which are identified as pions with opposite charge and the other two as either an $e^+e^-$ or $\mu^+\mu^-$ pair. Pion candidates must not be identified as electrons. To reject Bhabha events with bremsstrahlung followed by $\gamma$ conversions, we require the cosine of the polar angle of the electron with respect to the $e^-$ beam direction to satisfy $\cos\theta_{e^-} < 0.8$ in the laboratory frame. To improve the $e^\pm$ energy measurements, up to three photons are combined with $e^\pm$ candidates to partially recover bremsstrahlung [11]. The $\Upsilon(1S)$ candidate is selected by requiring $-0.35 < m_{e^+e^-} - m_{\Upsilon(1S)} < 0.2$ GeV/$c^2$ or $|m_{\mu^+\mu^-} - m_{\Upsilon(1S)}| < 0.2$ GeV/c$^2$, where the invariant mass of the lepton pair $m_{\ell^+\ell^-}$ is then constrained to the nominal $\Upsilon(1S)$ mass value [12]. The pion pair is combined with the $\Upsilon(1S)$ candidate to form a $\Upsilon(1^3D_J)$ candidate (mass resolution 3 MeV/$c^2$). To eliminate background from $\gamma \to e^+e^-$ conversions in which both the $e^+$ and $e^-$ are misidentified as pions, we reject events with a cosine for the laboratory $\pi^+\pi^-$ opening angle $\cos\theta_{\pi^+,\pi^-}$ greater than 0.95 if the converted $e^+e^-$ mass is less than 50 MeV/$c^2$, and events with a laboratory angle between the $\pi^+\pi^-$ pair and $\ell^\pm$ that satisfies $\cos\theta_{\pi^+\pi^-,\ell^\pm} > 0.98$.

Photons from $\Upsilon(3S) \to \gamma\chi_{bJ'}(2P)$ ($\chi_{bJ'}(2P) \to \gamma\Upsilon(1^3D_J)$) decays have energies between 86 and 122 MeV [12] (80 and 117 MeV [2]) in the $\Upsilon(3S)$ center-of-mass (CM) frame. Our resolution for 80 MeV photons is about 6.6 MeV. We require at least two photons in an event: one (the other) with CM energy larger than 70 MeV (60 MeV). Photons from final-state radiation (FSR) are rejected by requiring the cosines of the laboratory angles between the cascade photons and leptons to satisfy $\cos\theta_{\ell,\gamma} < 0.98$. In case of multiple photon combinations, we choose the one that minimizes $\chi^2 = \sum_{i=1,2}\left(E^i_\gamma - E^i_{\text{exp}}\right)^2/\sigma^2_{E^i_\gamma}$, where $E^i_{\text{exp}}$ are the nominal [12] (for $\Upsilon(3S) \to \gamma\chi_{bJ'}(2P)$) or expected [2] (for $\chi_{bJ'}(2P) \to \gamma\Upsilon(1^3D_J)$) photon energies that correspond to one of the six possible $\Upsilon(3S) \to \gamma\chi_{bJ'}(2P) \to$ $\gamma\gamma\Upsilon(1^3D_J)$ transition paths allowed by angular momentum conservation, with $E^i_\gamma$ ($\sigma_{E^i_\gamma}$) the measured energies (resolutions). We verified that the $\chi^2$ procedure does not bias our results, using simulated data samples in which the assumed $\Upsilon(1^3D_J)$ mass values are varied.

The $\Upsilon(1^3D_J)$ candidate is combined with the two photons to form a $\Upsilon(3S)$ candidate, whose CM momentum must be less than 0.3 GeV/$c$. The $\Upsilon(3S)$ mass is then constrained to its nominal value [12]. The $\Upsilon(3S)$ laboratory energy (resolution 25 MeV) is required to equal the summed $e^+$ and $e^-$ beam energies to within 0.1 GeV.

We identify four background categories within our fit interval $10.11 < m_{\pi^+\pi^-\ell^+\ell^-} < 10.28$ GeV/$c^2$: $\Upsilon(3S)$ decays to (I) $\gamma\chi_{bJ'}(2P)$ with $\chi_{bJ'}(2P) \to \omega\Upsilon(1S)$ and $\omega \to \pi^+\pi^-(\pi^0)$, (II) $\pi^+\pi^-\Upsilon(1S)$ with FSR, (III) $\eta\Upsilon(1S)$ with $\eta \to \pi^+\pi^-\pi^0(\gamma)$, and (IV) $\gamma\gamma\Upsilon(2S)$ or $\pi^0\pi^0\Upsilon(2S)$ with $\Upsilon(2S) \to \pi^+\pi^-\Upsilon(1S)$. Categories I and II are the main backgrounds.

An extended unbinned maximum likelihood fit is applied to the sample of 263 events in the fit interval. The fit has a component for each of the three $\Upsilon(1^3D_J)$ signal states and four background categories. The likelihood function is $\mathcal{L} = \exp\left(-\sum_j n_j\right)\prod_{i=1}^N\left[\sum_j n_j\mathcal{P}_j\left(m_i\right)\right]$, with $N$ the number of events, $n_j$ the yield of component $j$, $\mathcal{P}_j$ the probability density function (PDF) for component $j$, and $m$ the $\pi^+\pi^-\ell^+\ell^-$ invariant mass.

The PDFs are derived from MC simulations. Each signal PDF is parameterized by the sum of two Gaussians and a Crystal Ball (CB) function [13]. For background category I, we use the sum of a CB function, which describes the $\omega \to \pi^+\pi^-\pi^0$ events, and two Gaussians, which model the two peaks from $\chi_{b1,2}(2P)$ decays to $\omega\Upsilon(1S)$ with $\omega \to \pi^+\pi^-$. A bifurcated Gaussian, a high statistics histogram, and a Gaussian, model the PDFs for backgrounds II, III, and IV, respectively.

A large data control sample of $\Upsilon(3S) \to \gamma\chi_{bJ'}(2P) \to \gamma\gamma\Upsilon(2S)$ events with $\Upsilon(2S) \to \pi^+\pi^-\Upsilon(1S)$ and $\Upsilon(1S) \to \ell^+\ell^-$ is used to validate the signal PDFs and mass reconstruction. The control sample is selected using similar criteria to those used to select the $\Upsilon(1^3D_J)$. The background contamination is about 2%. Only a small difference is observed between the shapes of the $\Upsilon(2S) \to \pi^+\pi^-\ell^+\ell^-$ invariant mass distributions in data and simulation. The signal PDF is adjusted to account for this difference. The reconstructed $\Upsilon(2S)$ mass is shifted downwards by $0.70\pm0.15$ (stat.) MeV/$c^2$ compared to its nominal value [12]. We apply this shift as a correction to the $\Upsilon(1^3D_J)$ mass results presented below.

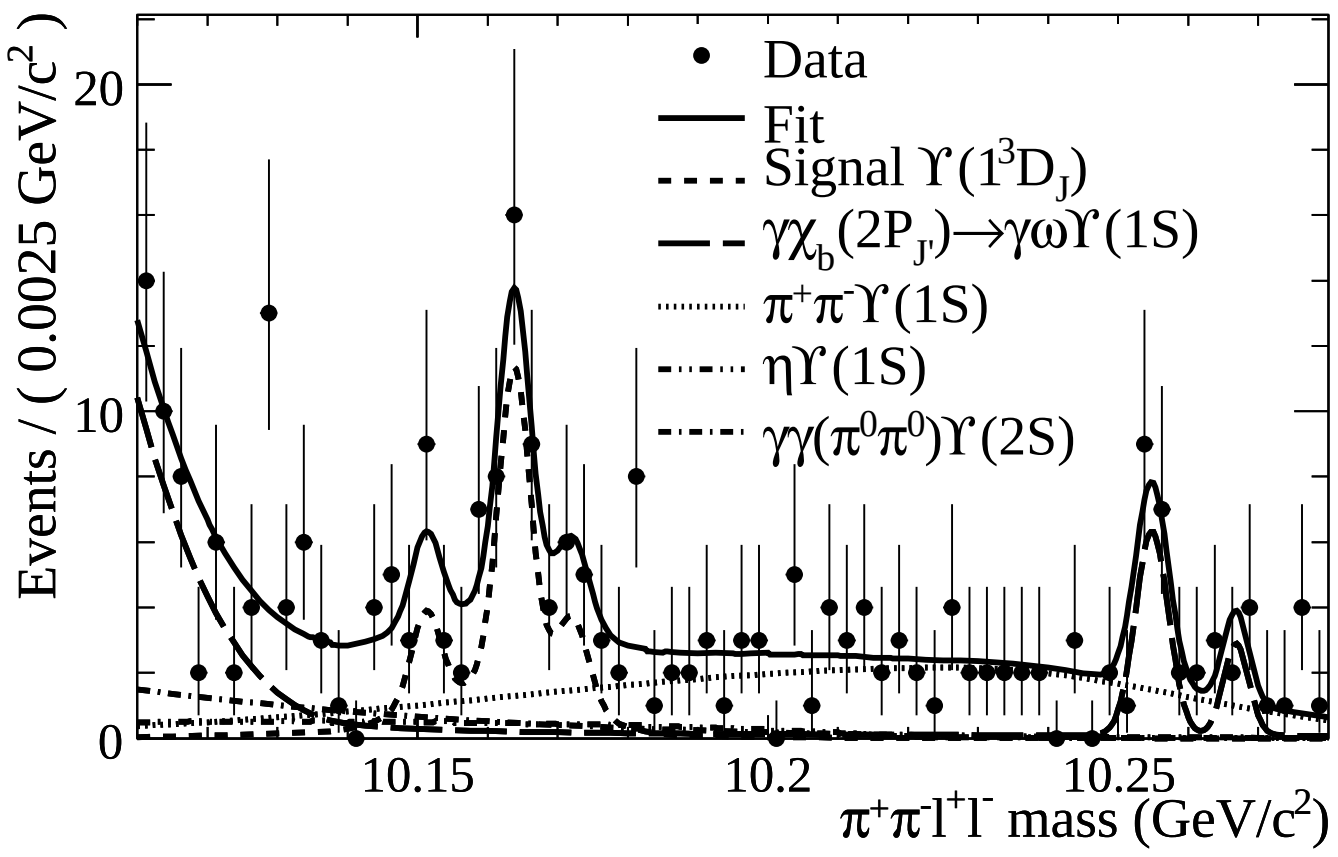

FIG. 1: The $\pi^+\pi^-\ell^+\ell^-$ mass spectrum and fit results. The two peaks near 10.25 GeV/$c^2$ arise from $\chi_{bJ'}(2P) \to \omega\Upsilon(1S)$ background events with $\omega \to \pi^+\pi^-$.

Eleven parameters are determined in the fit: the three signal yields and three masses, the yields of background categories I and II, and – within background category I – the $\chi_{b1}(2P)$ mass and the relative yields of the $\chi_{b1}(2P)$ and $\chi_{b2}(2P)$ peaks from $\omega \to \pi^+\pi^-$ decays. The mass difference between the $\chi_{b1}(2P)$ and $\chi_{b2}(2P)$ peaks is fixed to its measured value [12]. The yields of background categories III and IV are fixed to their expected values based on the measured branching fractions [12, 14].

Figure 1 shows the $\pi^+\pi^-\ell^+\ell^-$ mass distribution and fit results. The results for the separated $\Upsilon(1S) \to e^+e^-$ and $\Upsilon(1S) \to \mu^+\mu^-$ channels are shown in Fig. 2. The $e^+e^-$ channel has a smaller efficiency than the $\mu^+\mu^-$ channel in part because of the criteria to reject Bhabha events. The differences in efficiency between the $e^+e^-$ and $\mu^+\mu^-$ channels, including those for the $\chi_{bJ'}(2P) \to \omega\Upsilon(1S)$ background events, are consistent with the expectations from the simulation within the uncertainties. We find $10.6^{+5.7}_{-4.9}$ $\Upsilon(1^3D_1)$, $33.9^{+8.2}_{-7.5}$ $\Upsilon(1^3D_2)$, and $9.4^{+6.2}_{-5.2}$ $\Upsilon(1^3D_3)$ events. The positions of the three signal peaks in Fig. 1 are stable with respect to different initial assumptions about their masses within the fit interval. The fluctuations at around 10.13 and 10.18 GeV/$c^2$ are discussed below. The fitted background category I and II yields of $50\pm9$ and $94\pm13$ events agree with the MC expectations of 51 and 94 events, respectively. The fitted $\chi_{b1}(2P)$ mass value of $10255.7\pm0.7$ (stat.) MeV/$c^2$ (after applying the shift of $+0.7$ MeV/$c^2$ from the $\Upsilon(2S)$ mass calibration) is in good agreement with the nominal value $10255.5\pm0.5$ MeV/$c^2$ [12], validating the calibration.

Fit biases are evaluated by applying the fit to an ensemble of 2000 simulated experiments constructed by randomly extracting events from MC samples. The numbers of signal and background events and the $\Upsilon(1^3D_J)$ masses correspond to those of the fit. The biases are $1.6\pm0.1$, $-1.8\pm0.2$, and $1.0\pm0.1$ events for the $\Upsilon(1^3D_1)$, $\Upsilon(1^3D_2)$, and $\Upsilon(1^3D_3)$, respectively. We subtract these biases from the signal yields. The biases on the masses are negligible.

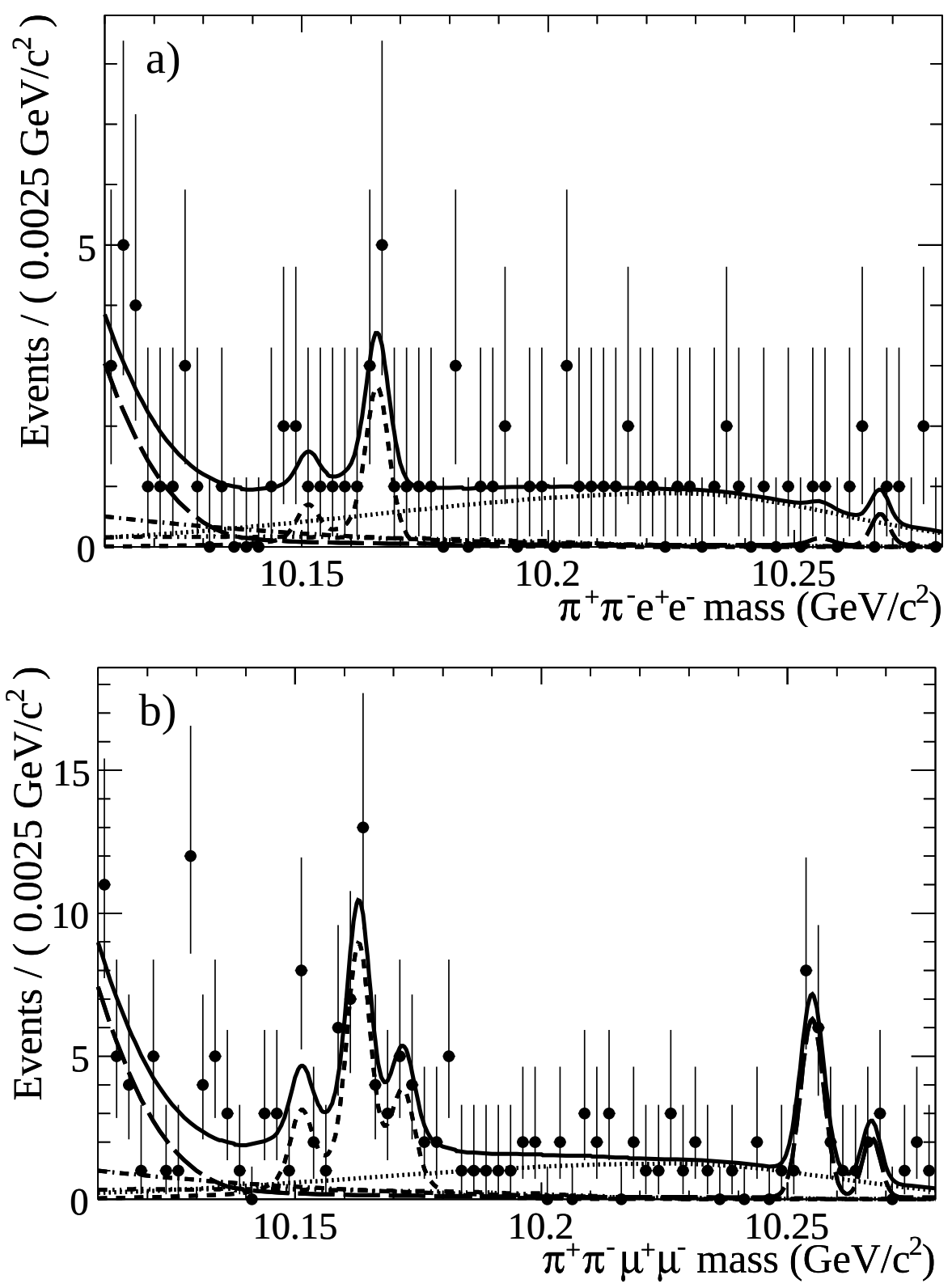

FIG. 2: The $\pi^+\pi^-\ell^+\ell^-$ mass spectra for the separated (a) $\Upsilon(1S) \to e^+e^-$ and (b) $\Upsilon(1S) \to \mu^+\mu^-$ channels. The results of the fit are shown. The legend is given in Fig. 1.

Multiplicative systematic uncertainties arise from the uncertainty in the number $N_{\Upsilon(3S)}$ of $\Upsilon(3S)$ events in the initial sample (1.0%) and in the reconstruction efficiencies for tracks (1.4%), photons (3.0%), and particle identification (2.0%). Additive systematic uncertainties originate from signal and background PDFs, evaluated by varying the PDF parameters within their uncertainties, background yields, evaluated by varying the background category IV (III) yield by its uncertainties (by $\pm100\%$), the fit bias, and the $\Upsilon(2S)$ mass calibration. The fit bias uncertainties are defined as the quadratic sum of half the biases and their statistical uncertainties. The mass calibration uncertainty is taken to be half the $\Upsilon(2S)$ mass shift added in quadrature with the $\Upsilon(2S)$ mass uncertainty [12]. The overall additive uncertainties for the signal yields (masses) are $1.5-2.0$ events ($0.48$ MeV/$c^2$) and are dominated by the contribution from the background yields ($\Upsilon(2S)$ mass calibration).

As a check, we repeat the fit with an additional background term, given by a second order polynomial. The purpose of this check is to test for the effect of potential unmodeled background. The parameters of the polynomial are left free in the fit (thus there are 14 free parameters). The fitted $\Upsilon(1^3D_J)$ yields are affected by less than 0.5 events compared to our standard fit, for all $J$ values. The shifts in the fitted mass values are less than 0.05 MeV. Since this polynomial background term is not motivated by any known source and since the description of the background without the additional term is good, we do not use this alternate background model to define a systematic uncertainty.

We define the statistical significance of each $\Upsilon(1^3D_J)$ state by the square root of the difference between the value of $-2\ln\mathcal{L}$ for zero signal events and assuming the bias-corrected signal yield, with the masses and yields of the other two states held at their fitted values. These results are validated with frequentist techniques. Systematics are included by convoluting $\mathcal{L}$ with a Gaussian whose standard deviation ($\sigma$) equals the total systematic uncertainty. The significances of the $\Upsilon(1^3D_1)$, $\Upsilon(1^3D_2)$, and $\Upsilon(1^3D_3)$ observations are 2.0 (1.8), 6.5 (5.8), and 1.7 (1.6) $\sigma$ without (with) systematics included, respectively. If we use the raw signal yields, rather than the bias-corrected yields, the statistical significances of the $J=1$, 2 and 3 states are 2.4, 6.2, and 2.0 $\sigma$, respectively.

From Fig. 1 it is seen that the data exhibit upward fluctuations at $\pi^+\pi^-\ell^+\ell^-$ masses around 10.13 and 10.18 GeV/$c^2$. To investigate the significance of these fluctuations, we re-perform the fit with the $J=1$ mass constrained to 10.13 GeV/$c^2$ rather than leaving it as a free parameter. An analogous fit is made with the $J=3$ mass constrained to 10.18 GeV/$c^2$. The statistical significance for this alternate $J=1$ ($J=3$) peak, evaluated using the raw signal yield, is 2.0 $\sigma$ (1.3 $\sigma$), compared to 2.4 $\sigma$ (2.0 $\sigma$) for our standard fit. The $J=2$ signal yield and mass shift by less than 1 event and 0.04 MeV/$c^2$, respectively, in these alternate fits.

We determine branching fractions by dividing the bias-corrected signal yields by the selection efficiencies and $N_{\Upsilon(3S)}$. The significances of the $\Upsilon(1^3D_1)$ and $\Upsilon(1^3D_3)$ peaks are low and we do not have clear evidence for them. For the $J=1$ and 3 states, we also present upper limits on the branching fractions assuming the fitted masses. The efficiencies for the six allowed $\Upsilon(3S) \to \gamma\chi_{bJ'}(2P) \to \gamma\gamma\Upsilon(1^3D_J)$ paths differ by up to 7.5% and therefore do not factorize, leaving six unknown branching fractions but only three measured signal yields. However, 91.4% of the $\Upsilon(3S) \to \gamma\gamma\Upsilon(1^3D_1)$ and 88.7% of the $\Upsilon(3S) \to \gamma\gamma\Upsilon(1^3D_2)$ transitions are predicted [2] to proceed through the $\chi_{b1}(2P)$ state, while $\Upsilon(3S) \to \gamma\gamma\Upsilon(1^3D_3)$ transitions can only proceed through the $\chi_{b2}(2P)$. Therefore, we evaluate the branching fractions for the dominant modes only, using the predicted ratios of the branching fractions to account for the non-dominant transitions. The efficiencies of the dominant modes, averaged over the $\Upsilon(1S) \to e^+e^-$ and $\mu^+\mu^-$ final states, are $26.7 \pm 0.1\%$, $26.7 \pm 0.1\%$, and $25.7 \pm 0.2\%$ for the $\Upsilon(1^3D_1)$, $\Upsilon(1^3D_2)$, and $\Upsilon(1^3D_3)$, respectively.

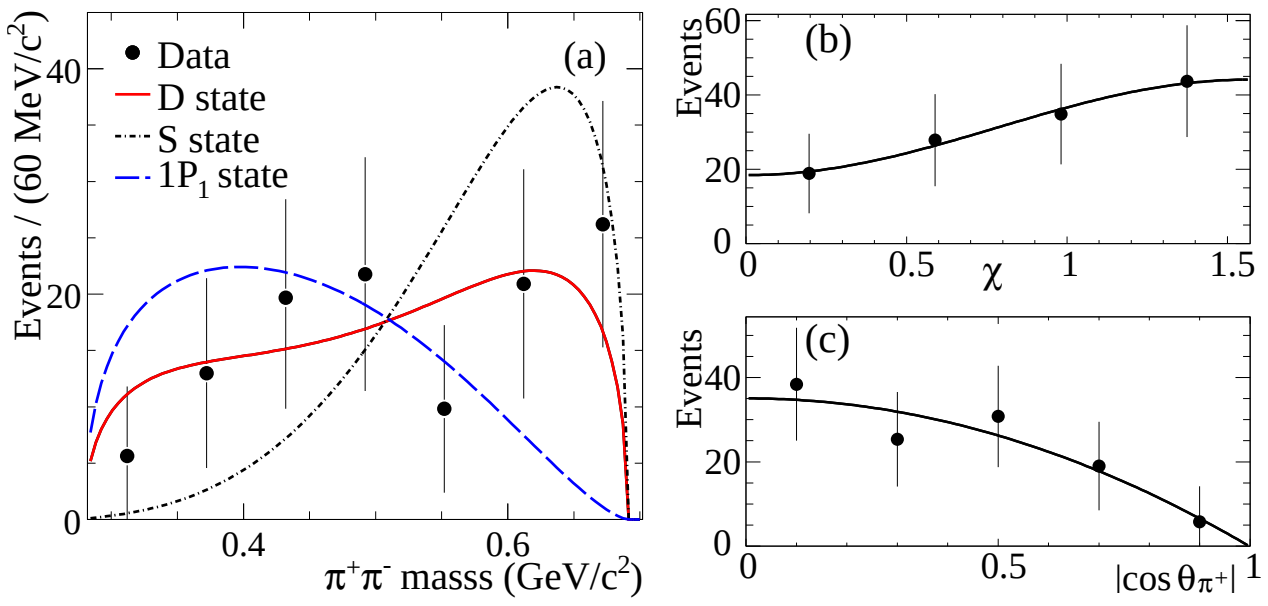

FIG. 3: (a) The $\pi^+\pi^-$ mass spectrum in the $\Upsilon(1^3D_2)$ signal region. The area under each curve equals the number of events. (b,c) Distributions in the $\Upsilon(1^3D_2)$ signal region of (b) the angle $\chi$ between the $\pi^+\pi^-$ and $\ell^+\ell^-$ planes, and (c) the $\pi^+$ helicity angle. The uncertainties include both statistical and systematic terms.

The branching fraction products for the dominant modes $\mathcal{B}_{J'J} \equiv \mathcal{B}_{\Upsilon(3S)\to\gamma\chi_{bJ'}(2P)} \times \mathcal{B}_{\chi_{bJ'}(2P)\to\gamma\Upsilon(1^3D_J)} \times \mathcal{B}_{\Upsilon(1^3D_J)\to\pi\pi\Upsilon(1S)} \times \mathcal{B}_{\Upsilon(1S)\to\ell\ell}$ (or the upper limits at 90% CL with systematics included) are, in units of $10^{-7}$, $\mathcal{B}_{11} = 1.27^{+0.81}_{-0.69} \pm 0.28\,(< 2.50)$, $\mathcal{B}_{12} = 4.9^{+1.1}_{-1.0} \pm 0.3$, and $\mathcal{B}_{23} = 1.34^{+0.99}_{-0.83} \pm 0.24\,(< 2.80)$. We determine the $\Upsilon(1^3D_2)$ mass to be $10164.5 \pm 0.8 \pm 0.5$ MeV/$c^2$, which is consistent with, and more precise than, the result $10\,161.1 \pm 0.6\,(\text{stat.}) \pm 1.6\,(\text{syst.})$ MeV/$c^2$ from CLEO [4].

From the $\Upsilon(3S) \to \gamma\chi_{bJ'}(2P)$ branching fractions and uncertainties [12] and $\chi_{bJ'}(2P) \to \gamma\Upsilon(1^3D_J)$ branching fraction predictions [2] we determine $\mathcal{B}[\Upsilon(1^3D_J) \to \pi^+\pi^-\Upsilon(1S)]$ (or 90% CL upper limits including systematics) to be $0.42\,^{+0.27}_{-0.23} \pm 0.10\%$ ($< 0.82\%$) for the $\Upsilon(1^3D_1)$, $0.66\,^{+0.15}_{-0.14} \pm 0.06\%$ for the $\Upsilon(1^3D_2)$, and $0.29\,^{+0.22}_{-0.18} \pm 0.06\%$ ($< 0.62\%$) for the $\Upsilon(1^3D_3)$, which lie between the predictions of about 0.2% from Ref. [7] and 2% from Ref. [8].

Figure 3(a) shows the $\pi^+\pi^-$ mass distribution for events in the $\Upsilon(1^3D_2)$ signal region $10.155 < m_{\pi^+\pi^-\ell^+\ell^-} < 10.168$ GeV/$c^2$ after subtraction of the backgrounds using the estimates from the fit. The data are corrected for mass-dependent efficiency variations. Shown in comparison are the expectations for the decay of a $D$ [15], $S$ [15], or ${}^1P_1$ [16] bottomonium state to $\pi^+\pi^-\Upsilon(1S)$. The resulting $\chi^2$ probabilities of 81%, 11%, and 10%, respectively, strongly favor the $D$ state.

The distribution of the angle $\chi$ between the $\ell^+\ell^-$ and $\pi^+\pi^-$ planes in the $\Upsilon(1^3D_J)$ rest frame, for events in the $\Upsilon(1^3D_2)$ signal region, is shown in Fig. 3(b). The data are corrected for background and efficiency. The $\chi$ distribution is expected to have the form $1 + \beta\cos 2\chi$ with $sign(\beta) = (-1)^J P$ [17], where $P$ is the parity. A fit to the data yields $\beta = -0.41 \pm 0.29\,(\text{stat.}) \pm 0.10\,(\text{syst.})$, consistent with the expected assignments $J = 2$ and $P = -1$.

The background-subtracted, efficiency-corrected distribution of the helicity angle $\theta_\pi$, for events in the $\Upsilon(1^3D_2)$ signal region, is shown in Fig. 3(c), where $\theta_\pi$ is the angle of the $\pi^+$ in the $\pi^+\pi^-$ rest frame with respect to the boost from the $\Upsilon(1^3D_2)$ frame. For $D$-state decays to $\pi^+\pi^-\Upsilon(1S)$, $\theta_\pi$ follows a $1 + \frac{\xi}{2}(3\cos^2\theta_\pi - 1)$ distribution, where $\xi$ is a dynamical parameter to be determined experimentally. For $S$-state decays, the $\theta_\pi$ distribution is flat ($\xi = 0$). A fit to data yields $\xi = -1.0 \pm 0.4\,(\text{stat.}) \pm 0.1\,(\text{syst.})$, disfavoring the $S$ state.

In summary, we have observed the $\Upsilon(1^3D_2)$ bottomonium state through decays to $\pi^+\pi^-\Upsilon(1S)$. The significance is $5.8\sigma$ including systematic uncertainties. We improve the measurement of the $\Upsilon(1^3D_2)$ mass and determine the $\Upsilon(1^3D_J) \to \pi^+\pi^-\Upsilon(1S)$ branching fractions or set upper limits. We use the $\pi^+\pi^-$ invariant mass, the angle between the $\pi^+\pi^-$ and $\ell^+\ell^-$ planes, and the $\pi^+$ helicity angle, to test the consistency of the observed state with the expected quantum numbers $L = 2$, $J = 2$ and parity $P = -1$ for the dominant member of the triplet $\Upsilon(1^3D_2)$.

We are grateful for the excellent luminosity and machine conditions provided by our PEP-II colleagues, and for the substantial dedicated effort from the computing organizations that support *BABAR*. The collaborating institutions wish to thank SLAC for its support and kind hospitality. This work is supported by DOE and NSF (USA), NSERC (Canada), CEA and CNRS-IN2P3 (France), BMBF and DFG (Germany), INFN (Italy), FOM (The Netherlands), NFR (Norway), MES (Russia), MEC (Spain), and STFC (United Kingdom). Individuals have received support from the Marie Curie EIF (Euro-pean Union) and the A. P. Sloan Foundation.

---

* Now at Temple University, Philadelphia, Pennsylvania 19122, USA

† Also with Università di Perugia, Dipartimento di Fisica, Perugia, Italy

‡ Also with Università di Roma La Sapienza, I-00185 Roma, Italy

§ Now at University of South Alabama, Mobile, Alabama 36688, USA

¶ Also with Università di Sassari, Sassari, Italy